\newcommand{\be}{\begin{eqnarray}}
\newcommand{\ee}{\end{eqnarray}}
\documentclass[twocolumn,aps,showpacs]{revtex4}
\usepackage{graphics}
\begin{document}
\title{Pion dispersion relation at finite density and temperature}
\author{Alejandro Ayala}  
\affiliation{Instituto de Ciencias Nucleares, Universidad Nacional 
         Aut\'onoma de M\'exico, Apartado Postal 70-543, 
         M\'exico Distrito Federal 04510, M\'exico.}
\author{Paolo Amore}
\affiliation{Facultad de Ciencias, Universidad de Colima,
         Av. 25 de Julio 965, Colima Colima, 28045 M\'exico.}
\author{Alfredo Aranda}
\affiliation{Physics Department, Boston University, 
         590 Commonwealth Ave. Boston, MA 02215.}
\begin{abstract}

We study the behavior of the pion dispersion relation in a pion medium
at finite density and temperature. We introduce a pion chemical
potential to describe the finite pion number density and argue that such
description is valid during the hadronic phase of a relativistic
heavy-ion collision between chemical and thermal freeze-out. We make
use of an effective Lagrangian that explicitly respects chiral
symmetry through the enforcement of the chiral Ward identities. The pion
dispersion relation is computed through the computation of the pion
self-energy in a non-perturbative fashion by giving an approximate
solution to the Schwinger-Dyson equation for this self-energy. The
dispersion relation is described in terms of a density and temperature
dependent mass and an index of refraction which is also temperature, density 
as well as momentum dependent. The index of refraction is larger than
unity for all values of the momentum for finite $\mu$ and $T$. We
conclude by exploring some of the possible consequences for the
propagation of pions through the boundary between the medium and vacuum. 

\end{abstract}


\maketitle

\section{Introduction}

The phase structure of hadronic matter at high density and temperature
has become a subject of increasing attention in recent years, both
from the theoretical and experimental points of view. This
attention is driven mainly by the possibility to produce a locally
thermalized, deconfined state, whose degrees of
freedom are the quarks and gluons of quantum chromodynamics (QCD), the
so called quark-gluon plasma (QGP), in high-energy heavy-ion
collisions~\cite{QM01}.  

If the QGP is produced in these kind of reactions, the prevailing view
portraits the evolution of such a system traversing a series of
stages, the last of which consists of a large amount of hadrons
strongly interacting in a finite volume, until a final freeze-out.

For not too high temperatures, hadronic matter consists mainly of
pions, therefore, the study of the propagation properties of pions
within the above described environment represents an important
ingredient for the understanding of the properties of the hadronic
system at and just before freeze-out. For instance, it has been 
speculated that the pion interactions within a dense hadronic
environment can give rise to interesting collective surface 
phenomena~\cite{Shuryak}. Moreover, it is also well known that the pion group 
velocity is an important piece of information necessary to properly account 
for the dilepton spectrum coming from the hadronic phase~\cite{Gale-Liu}. 

The hadronic degrees of freedom are customarily accounted for by means
of effective chiral theories that incorporate the Goldstone boson
nature of pions. One of such theories is chiral perturbation theory
(ChPT) which has been employed to show the well known result that at
leading perturbative order and at low momentum, the modification of
the pion dispersion curve in a pion medium at finite temperature is just
a constant, temperature dependent, increase of the pion
mass~\cite{Gasser}. ChPT has also been used in a two-loop computation
of the pion self-energy~\cite{Schenk} and decay
constant~\cite{Toublan}. A striking result obtained from such
computations is that at second order, the shift in the temperature
dependence of the pion mass is opposite in sign and about three times
larger in magnitude than the first order shift, already at
temperatures close to $150$ MeV. This result might signal either the
breakdown of the perturbative expansion at these temperatures or the need
to perform such calculations using schemes other than perturbation theory.  

However, a missing ingredient in the calculations of the pion
dispersion curve is the treatment of the medium's finite density. The
conceptual difficulty is related to the fact that, though it is
possible to assume that the system is in (at least local) thermal
equilibrium, strictly speaking the only conserved charge that can
be associated to the pion system is the electric charge and thus, for
an electrically neutral pion system the corresponding chemical potential 
vanishes. The behavior of the pion mass in the presence of an isospin
chemical potential has been recently studied in Ref.~\cite{Loewe}. But
in order to describe a situation 
in which the number of pions in thermal equilibrium is finite, we need to
consider a chemical potential associated with the
pion number, instead of its charge. 

Recall that the pion number is not a conserved quantity due to either strong,
weak or electromagnetic processes. Nevertheless, the characteristic
time for electromagnetic and weak pion number-changing processes, is
very large compared to the lifetime of the system created in
relativistic heavy-ion collisions and therefore, these processes are of no
relevance for the propagation properties of pions within the hadronic
phase of the collision. As for the case of strong
processes, it is by now accepted that they drive pion number toward
chemical freeze-out at a temperature considerably higher than the
thermal freeze-out temperature and therefore, that
from chemical to thermal freeze-out, the pion system evolves with the
pion abundance held fixed~\cite{Bebie, Braun-Munzinger}. Under these
circumstances, it is possible to ascribe to the pion density a
chemical potential and consider the pion number as
conserved~\cite{Hung,Chungsik}. In this context, the role of a finite
pion chemical potential into a hadronic equation of state has been recently
investigated in Ref.~\cite{Teaney}.

Furthermore, another important ingredient in the analysis is the well
know fact that in finite temperature field theories with either
massless degrees of freedom or that exhibit spontaneous symmetry
breaking~\cite{Dolan}, the perturbative expansion breaks down and thus
the necessity to implement resummation techniques. 

In this paper we explore the effects introduced by a finite pion
density on the dispersion curve of pions at finite
temperature. Starting from the linear sigma model, we use an effective
Lagrangian~\cite{Ayala} obtained by integrating out the heavy sigma modes and
compute, in a non perturbative fashion, the pion self-energy. We find
that the dispersion curve is modified with respect to the vacuum in a
way described by the introduction of an index of refraction larger
than one, in addition to the thermal and density increase of the pion
mass. We discuss possible implications of such behavior of the 
pion dispersion curve. 

The work is organized as follows: In section~\ref{secII}, we recall
how for temperatures and momenta smaller than the mass of the sigma,
it is possible to integrate out the heavy sigma modes to construct an
effective Lagrangian that reproduces the two-loop pion self-energy at
finite temperature. In section~\ref{secIII}, we use this 
effective Lagrangian to compute non-perturbatively the pion
self-energy and from it, the pion dispersion curve at finite
density and temperature. Finally, in section~\ref{secIV}, we summarize
and discuss our results, emphasizing some of the possible consequences
of the behavior of the dispersion curve thus found for the propagation
pions that approach the boundary between the hadronic medium and
vacuum in a relativistic heavy-ion collision.  

\section{Effective Lagrangian}\label{secII}

The Lagrangian for the linear sigma model, including only meson degrees of 
freedom and with an explicit chiral symmetry breaking term, 
can be written as~\cite{Lee}
\be
   {\mathcal{L}}&=&\frac{1}{2}\left[(\partial{\mathbf{\pi}})^2 +
                (\partial\sigma)^2 - m_\pi^2{\mathbf{\pi}}^2 - 
                m_\sigma^2\sigma^2\right]\nonumber\\ 
                &-&\lambda^2 f_\pi\sigma (\sigma^2 + {\mathbf{\pi}}^2) -
                \frac{\lambda^2}{4}(\sigma^2 + {\mathbf{\pi}}^2)^2\, ,
   \label{lagrangian}
\ee
where $\mathbf{\pi}$ and  $\sigma$ are the pion and sigma fields,
respectively, and the coupling $\lambda^2$ is given by
\be
   \lambda^2=\frac{m_\sigma^2-m_\pi^2}{2f_\pi^2}\, .
   \label{coupling}
\ee

From the above Lagrangian one obtains the Green's functions and Feynman 
rules to be used in perturbative calculations, in the usual
manner. For instance, the bare pion and sigma propagators 
$\Delta_\pi (P)$, $\Delta_\sigma (Q)$ and the bare one-sigma two-pion and 
four-pion vertices $\Gamma_{12}^{ij}$, $\Gamma_{04}^{ijkl}$ are given by 
(hereafter, capital Roman letters are used to denote four momenta)
\be
   i\Delta_\pi(P)\delta^{ij}&=&\frac{i}{P^2-m_\pi^2}\delta^{ij}\nonumber \\
   i\Delta_\sigma (Q)&=&\frac{i}{Q^2-m_\sigma^2}\nonumber \\
   i\Gamma_{12}^{ij}&=&-2i\lambda^2 f_\pi\delta^{ij}\nonumber \\
   i\Gamma_{04}^{ijkl}&=&-2i\lambda^2(\delta^{ij}\delta^{kl} + 
                          \delta^{ik}\delta^{jl} + \delta^{il}\delta^{jk})\, .
   \label{rules}
\ee
These Green's functions are sufficient to obtain the modification to the 
pion propagator, both at zero and finite temperature, at any given 
perturbative order. 

\begin{figure} 
{\centering
\resizebox*{0.2\textwidth}{0.05\textheight}
{\includegraphics{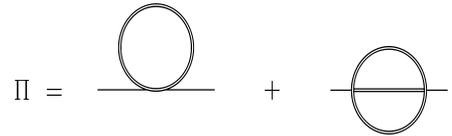}}\par} 
\caption{Diagramatic representation of the self-consistent pion
self-energy. The double lines represent the full pion
propagator.}
\end{figure}

When interested in a given approximation scheme to build such Green's
functions, it is possible to exploit the relations that chiral 
symmetry imposes among them. These relations, better known as chiral
Ward identities (ChWI), are a direct consequence of the fact that the
divergence of the axial current may be used 
as an interpolating field for the pion~\cite{Lee}. Thus, one can construct
the modification to one of the above Green's functions at a given perturbative
order and for a given approximation. In order to make sure that the
approximation respects chiral symmetry, one needs to check that the
modification of other Green's functions respect the corresponding
ChWI. For example, two of the ChWI satisfied --order by order in
perturbation theory-- by the functions  
$\Delta_\pi (P)$, $\Delta_\sigma (Q)$, $\Gamma_{12}^{ij}$ and 
$\Gamma_{04}^{ijkl}$ are
\be
   f_\pi\Gamma_{04}^{ijkl}(;0,P_1,P_2,P_3)&=&
   \Gamma_{12}^{kl}(P_1;P_2,P_3)\delta^{ij}\nonumber\\ 
   &+& 
   \Gamma_{12}^{lj}(P_2;P_3,P_1)\delta^{ik}\nonumber\\ 
   &+&
   \Gamma_{12}^{jk}(P_3;P_1,P_2)\delta^{il}\nonumber\\
   f_\pi\Gamma_{12}^{ij}(Q;0,P)&=&
   \left[\Delta_\sigma^{-1}(Q) - 
   \Delta_\pi^{-1}(P)\right]\delta^{ij}
   \label{Ward}
\ee
where momentum conservation at the vertices is implied, that is
$P_1+P_2+P_3=0$ and $Q+P=0$. The notation for the functional
dependence of the vertices in Eqs.~(\ref{Ward}) is such that the
variables before and after the semicolon refer to the four-momenta of
the sigma and pion fields, respectively~\cite{Lee}.  

In Refs.~\cite{Ayala} it has been shown that in the kinematical regime
where the pion momentum, the pion mass and the temperature are small compared
to the sigma mass, the effective one-loop sigma propagator
and one-sigma two-pion and four-pion vertices are given by 
\be
   i\Delta_\sigma^\star (Q)&=&\frac{i}{Q^2 - m_\sigma^2 + 
                              6\lambda^4f_\pi^2{\mathcal I}^t
                              (Q) }\, .
   \label{newsigprop}
\ee
\be
   i\Gamma_{12}^{\star\, ij}(Q;P_1,P_2)=
   -2i\lambda^2 f_\pi\delta^{ij}
   \left[1 - 3\lambda^2{\mathcal I}^t(Q)\right]\, ,  
   \label{vert12tot}
\ee
\be
   i\Gamma_{04}^{\star\ ijkl}(;P_1,P_2,P_3,P_4)\!\!&=&\!\!
   2i\lambda^2\left\{\right.
   \nonumber\\
   \!\!&\times&\!\!
   \left.\left[1 - 3\lambda^2{\mathcal I}^t(P_1+P_2)\right]
   \delta^{ij}\delta^{kl}\right.\nonumber\\
   \!\!&+&\!\! 
   \left[1 - 3\lambda^2{\mathcal I}^t(P_1+P_3)\right]
   \delta^{ik}\delta^{jl}\nonumber\\
   \!\!&+&\!\! \left.
   \left[1 - 3\lambda^2{\mathcal I}^t(P_1+P_4)\right]\delta^{il}
   \delta^{jk}\right\},\nonumber\\
   \label{vert4tot}
\ee
where in the imaginary-time formalism of thermal field theory (TFT), the 
function ${\mathcal I}^t$ is obtained as the time-ordered analytical
continuation to real energies of the function ${\mathcal I}$ defined by
\be
   {\mathcal I}(Q)&\equiv& T\sum_n\int\frac{d^3k}{(2\pi)^3}
   \nonumber\\
   &\times&
   \frac{1}{K^2+m_\pi^2}
   \,\frac{1}{(K-Q)^2+m_\pi^2}\, .
   \label{funcIJ}
\ee
Here $Q=(\omega,{\mathbf{q}}),\, K=(\omega_n,{\mathbf{k}})$,
$Q^2=\omega^2 + q^2$, $K^2=\omega_n^2 + k^2$ with
$\omega = 2m\pi T$ and $\omega_n = 2n\pi T$ ($m$, $n$ integers) being
discrete boson frequencies, $T$ is the temperature and
$q=|{\mathbf{q}}|,\, k=|{\mathbf{k}}|$. 

It is easy to check that Eqs.~(\ref{newsigprop}), (\ref{vert12tot}) and 
(\ref{vert4tot}) satisfy the Ward identities in Eq.~(\ref{Ward}), this ensures
that the approximation scheme adopted respects chiral symmetry.

By using the above effective vertices and propagator, it is possible
to construct the two-loop modification to the pion self-energy in the
same kinematical regime with the result~\cite{Ayala}
\be
   \Pi_2(P)&=&\left(\frac{m_\pi^2}{2f_\pi^2}\right)
   T\sum_n\int\frac{d^3k}{(2\pi)^3}\frac{1}{K^2+m_\pi^2}\nonumber\\
   &\times&
   \left\{3-\left(\frac{m_\pi^2}{2f_\pi^2}\right)
   \left[9{\mathcal I}^t(0) + 6{\mathcal I}^t(P+K)\right]\right\}.
   \label{selfmod}
\ee
Equation~(\ref{selfmod}) reproduces the leading order result obtained from
ChPT~\cite{Gasser}. Furthermore, we observe that Eq.~(\ref{selfmod})
can be formally obtained by means of the effective Lagrangian 
\be
   {\mathcal L}=\frac{1}{2}\left(\partial_\mu{\mathbf{\phi}}\right)^2
   -\frac{1}{2}m_\pi^2{\mathbf{\phi}}^2 -\frac{\alpha}{4!}\left(
   {\mathbf{\phi}}^2\right)^2\, ,
   \label{effLag}
\ee
where $\alpha=6(m_\pi^2/2f_\pi^2)$ and the factor $6$ comes from
considering the interaction of like-isospin pions in the vertex
\be
   i\Gamma_4^{ijkl}=-2i\left(\frac{m_\pi^2}{2f_\pi^2}\right)
   \left(\delta^{ij}\delta^{kl}+\delta^{ik}\delta^{jl}+\delta^{il}\delta^{jk}
   \right)\, .
   \label{vertmod}
\ee
Eqs.~(\ref{selfmod}) and~(\ref{effLag}) mean that in the kinematical
regime where the sigma mass is large compared to the pion mass,
the momentum and the temperature, the linear sigma model Lagrangian
reduces to a $\phi^4$ Lagrangian for effective like-isospin pions with
an effective coupling given by $\alpha$. By restricting ourselves to
the above kinematical regime, we will proceed on working with the
Lagrangian given by Eq.~(\ref{effLag}). 

\section{Non-perturbative pion self-energy}\label{secIII}

It is well know that quantum field theories at finite temperature
present certain subtleties such as the breakdown of the perturbative
expansion~\cite{Kapusta}. This breakdown becomes manifest in two
important cases: the appearance of infrared divergences in theories
with massless degrees of freedom, and the compensation of powers of the
coupling constant with powers of $T$ for large temperatures. In both
situations, the resummation of certain classes of diagrams represents
an important improvement for the study of the physical properties of such
theories. Even for cases where neither there is a massless degree of
freedom, nor the temperature is extremely large, it is important to
consider a resummation scheme, particularly for the case of theories with
spontaneous symmetry breaking near critical behavior.

In order to consider the above mentioned general situation and with the
purpose of studying the pion dispersion curve at finite
density and temperature, let us formally consider the
Schwinger-Dyson equation for the pion self-energy depicted in Fig.~1,
whose analytic expression is given by
\be
   \Pi(P)&=&\frac{\alpha}{2}T\sum_n\int\frac{d^3k}{(2\pi)^3}
   \frac{1}{K^2+m_\pi^2+\Pi}\nonumber\\
   &-&\frac{\alpha^2}{6}T^2\sum_{n_1,n_2}\int
   \frac{d^3k_1}{(2\pi)^3}\frac{d^3k_2}{(2\pi)^3}\frac{1}{K_1^2+m_\pi^2+\Pi}
   \nonumber\\
   &\times&
   \frac{1}{K_2^2+m_\pi^2+\Pi}
   \frac{1}{(K_1+K_2-P)^2+m_\pi^2+\Pi}\, ,\nonumber\\
   \label{SD}
\ee
where for internal lines, we make the substitution $i\omega\rightarrow
i\omega +\mu$.
Notice that since the interaction Lagrangian contains only a quartic
term, there is no need to dress the vertices in the above
equation.

Equation~(\ref{SD}) represents an integral equation for the function
$\Pi(P)$, which, needless to say, is very difficult to be solved
exactly. In order to find an approximate solution let us write 
\be
   \Pi(P)\simeq\Pi_0+\tilde{\Pi}(P)
   \label{decomp}
\ee
and consider $\tilde{\Pi}(P)\ll\Pi_0$. As we will see, such assumption
is justified given that in our approximation, $\tilde{\Pi}\sim{\mathcal
O}(\alpha^2)$. Keeping only the lowest order contribution in
$\tilde{\Pi}$, Eq.~(\ref{SD}) becomes
\be
   \Pi(P)&\equiv&\Pi_0 + \tilde{\Pi}(P)\nonumber\\
   &=&\frac{\alpha}{2}T\sum_n\int\frac{d^3k}{(2\pi)^3}
   \frac{1}{K^2+m_\pi^2+\Pi_0}\nonumber\\
   &-&\frac{\alpha^2}{6}T^2\sum_{n_1,n_2}\int
   \frac{d^3k_1}{(2\pi)^3}\frac{d^3k_2}{(2\pi)^3}\frac{1}{K_1^2+m_\pi^2+\Pi_0}
   \nonumber\\
   &\times&
   \frac{1}{K_2^2+m_\pi^2+\Pi_0}
   \frac{1}{(K_1+K_2-P)^2+m_\pi^2+\Pi_0}\, ,\nonumber\\
   \label{SDmod}
\ee
which in turn, serves as the definition of the functions $\Pi_0$ and
$\tilde{\Pi}(P)$, given by
\be
   \Pi_0\equiv\frac{\alpha}{2}T\sum_n\int\frac{d^3k}{(2\pi)^3}
   \frac{1}{K^2+m_\pi^2+\Pi_0}\, ,
   \label{pi0}
\ee
\be
   \tilde{\Pi}(P)&\equiv&-\frac{\alpha^2}{6}T^2\sum_{n_1,n_2}\int
   \frac{d^3k_1}{(2\pi)^3}\frac{d^3k_2}{(2\pi)^3}\frac{1}{K_1^2+m_\pi^2+\Pi_0}
   \nonumber\\
   &\times&
   \frac{1}{K_2^2+m_\pi^2+\Pi_0}
   \frac{1}{(K_1+K_2-P)^2+m_\pi^2+\Pi_0}\, .\nonumber\\
   \label{pitilde}
\ee
Equation~(\ref{pi0}) represents a self consistent equation for the
(momentum independent) constant $\Pi_0$. This is the well known
resummation for the {\it superdaisy} diagrams which constitute the dominant
contribution in the large-$N$ expansion~\cite{Dolan} of the Lagrangian in
Eq.~(\ref{effLag}). On the other hand, Eq.~(\ref{pitilde}) represents a first
approximation to the momentum dependent piece of $\Pi (P)$. 

The solution to Eq.~(\ref{pi0}) is given by the transcendental equation
\be
   \Pi_0&=&\left(\frac{\alpha T}{4\pi}\right)\sqrt{m_\pi^2 + \Pi_0}
   \nonumber\\
   &\times&\sum_{n=1}^\infty K_1\left(\frac{n\sqrt{m_\pi^2 + \Pi_0}}{T}\right)
   \frac{\cosh (n\mu /T)}{n}\, .
   \label{solpi0}
\ee
Figure~2 shows the behavior of the quantity $\sqrt{m_\pi^2 + \Pi_0}$
as a function of $(a)$ the temperature $T$ for different values of the
chemical potential $\mu$ and $(b)$ as a function of $\mu$ for
different values of $T$. In both cases, $\sqrt{m_\pi^2 + \Pi_0}$ grows
monotonically with both $T$ and $\mu$.

\begin{figure}
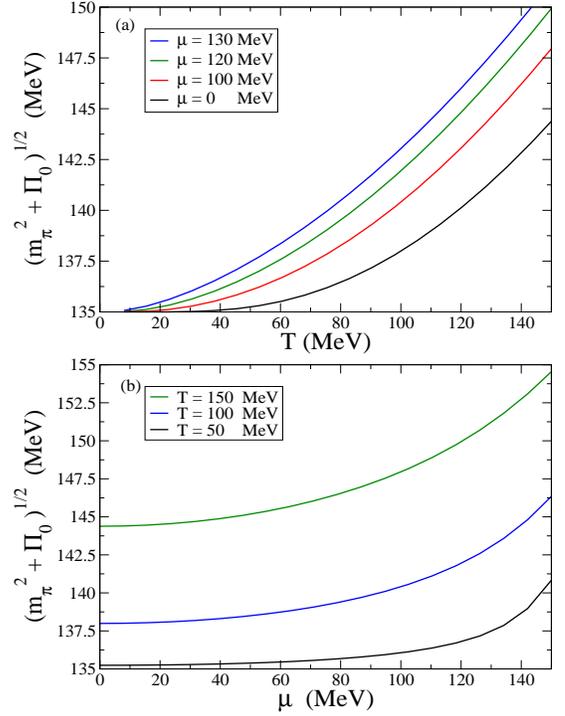
 
{\centering
\resizebox*{0.4\textwidth}{0.2\textheight}{\includegraphics{figure2a.eps}}
\par}
{\centering 
\resizebox*{0.4\textwidth}{0.2\textheight}{\includegraphics{figure2b.eps}}
\par}
\caption{$\sqrt{m_\pi^2 + \Pi_0}$ as a function of $(a)$ the
temperature $T$ for different values of the chemical potential ranging
from $\mu=0$ to $\mu=130$ MeV and as a function of $(b)$ the
chemical potential $\mu$ for different values of the temperature
ranging from $T=50$ to $T=150$ MeV.}
\end{figure}

The calculation of the function $\tilde{\Pi}(P)$ involves that of the
function
\be
   {\mathcal S}(P)\equiv
   T\sum_n\int\frac{d^3k}{(2\pi)^3}
   \frac{{\mathcal I}(K;\sqrt{m_\pi^2+\Pi_0})}
   {(K-P)^2+m_\pi^2+\Pi_0}\, ,
   \label{defS} 
\ee
where ${\mathcal I}(K;\sqrt{m_\pi^2+\Pi_0})$ is the function
defined in Eq.~(\ref{funcIJ}), with the replacement 
$m_\pi\rightarrow\sqrt{m_\pi^2+\Pi_0}$. The calculation
of the dispersion relation requires knowledge of the real part of the
retarded version of the function ${\mathcal S}(P)$, which, after analytical
continuation to real frequencies, is given
by~\cite{LeBellac} 
\be
   {\mbox R}{\mbox e}{\mathcal S}^r(p_0,p)\!\!\!&\equiv&\!\!\!\frac{1}{2}
   \left[{\mathcal S}(i\omega\rightarrow p_0+i\epsilon ,p)
   \right.\nonumber\\
   \!\!\!&+&\!\!\!\left. 
   {\mathcal S}(i\omega\rightarrow p_0-i\epsilon ,p)\right]\nonumber\\
   \!\!\!&=&\!\!\!\ -\ {\mathcal P}\int\frac{d^3k}{(2\pi)^3}
   \int_{-\infty}^{\infty}\frac{dk_0}{2\pi}\int_{-\infty}^{\infty}
   \frac{dk_0'}{2\pi}\nonumber\\
   \!\!\!&&\!\!\!
   \left[1+f(k_0+\mu)+f(k_0'-\mu)\right]
   2\ {\mbox I}{\mbox m}{\mathcal I}^t(k_0,k)\nonumber\\
   \!\!\!&&\!\!\!\frac{2\pi\ \varepsilon(k_0')\ 
   \delta[{k_0'}^2-({\mathbf{k}}-{\mathbf{p}})^2
   -m_\pi^2-\Pi_0]}{p_0-k_0-k_0'},\nonumber\\
   \label{reS}
\ee
where ${\mathcal P}$ stands for the principal part of the integral,
$\varepsilon$ is the sign function and 
\be
   f(x)=\frac{1}{e^{x/T}-1}
   \label{BE}
\ee 
is the Bose-Einstein distribution.

From Eq.~(\ref{reS}) we see that we require only knowledge of the
imaginary part of ${\mathcal I}^t(k_0,k)$. It is a lengthy, though
straightforward exercise to show that
\be
   {\mbox I}{\mbox m}{\mathcal I}^t(k_0,k)\!\!\!&=&\!\!\!
   -\frac{1}{16\pi}
   \Theta[K^2-4(m_\pi^2+\Pi_0)]\nonumber\\
   \!\!\!&\times&\!\!\!
   \{ [ a(K^2)+\frac{2T}{k}\ln\left(
   \frac{1-e^{-[\omega_+(k_0,k)+\mu] /T}}
   {1-e^{-[\omega_-(k_0,k)+\mu] /T}}\right)]
   \nonumber\\
   \!\!\!&\times&\!\!\!\Theta(k_0)\nonumber\\
   \!\!\!&+&\!\!\! [ a(K^2)+\frac{2T}{k}\ln\left(
   \frac{1-e^{-[\omega_+(k_0,k)-\mu] /T}}
   {1-e^{-[\omega_-(k_0,k)-\mu] /T}}\right)]
   \nonumber\\
   \!\!\!&\times&\!\!\!\Theta(-k_0)\}\, ,
   \label{imtimeorder}
\ee
where $K^2=k_0^2-k^2$, $\Theta$ is the step function and the functions 
$a$ and $\omega_\pm$ are 
\be
   a(K^2)&=&\sqrt{1-\frac{4(m_\pi^2+\Pi_0)}{K^2}}\nonumber \\
   \omega_\pm (k_0,k)&=&\frac{|k_0| \pm a(K^2)k}{2}\, .
   \label{imfunc}
\ee
It can also be checked that for $\mu=0$, Eqs.~(\ref{imtimeorder})
and~(\ref{imfunc}) reduce to the corresponding expressions found in
Refs.~\cite{Ayala}.

\begin{figure} 
{\centering
\resizebox*{0.4\textwidth}{0.2\textheight}{\includegraphics{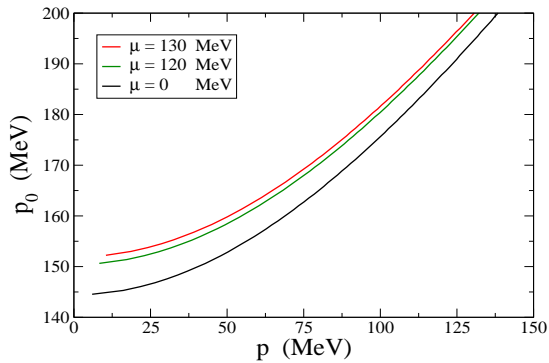}}
\par} 
\caption{$p_0$ as a function of $p$ for $\mu=0,120,130$ MeV and
$T=120$ MeV.} 
\end{figure}

Notice that Eq.~(\ref{reS}) contains temperature-dependent infinities 
coming from the terms involving the function ${\mathcal I}^t$, as well as 
vacuum infinities. This is an usual feature of multi-loop calculations
at finite temperature where one always encounters
temperature-dependent infinities in integrals involving only the bare
terms of the original Lagrangian. However,  
as it turns out, these infinities are exactly canceled by the contribution 
from the integrals computed by using the counterterms that are necessary to 
introduce at the previous order of the loop expansion to carry the
(vacuum) renormalization. This renormalization method has been
developed in Refs.~\cite{Caldas} for the resummation method employed
here and has been given the name {\it Modified Self-Consistent
Resummation}. For the purposes of this work, we concentrate on the
finite, temperature-dependent terms and refer the reader to the above
cited works for a detailed treatment of the renormalization
procedure.

According to Eqs.~(\ref{pitilde}) and~(\ref{defS})
\be
   {\mbox R}{\mbox e}\tilde{\Pi}^r(p_0,p)=-\frac{\alpha^2}{6}
   {\mbox R}{\mbox e}{\mathcal S}^r(p_0,p)\, ,
   \label{SandPi}
\ee
therefore, the pion dispersion relation is given as the solution to
\be
   p_0^2-\left[p^2+m_\pi^2+\Pi_0 - 
   \frac{\alpha^2}{6}
   {\mbox R}{\mbox e}{\mathcal S}^r(p_0,p)\right]=0
   \label{solution}
\ee
for positive $p_0$. Figure~3 shows plots of 
$p_0$ as a function of $p$ for
different values of $\mu$ and a temperature $T=120$ MeV. As can be
seen from this figure, $\Pi(p_0,p=0)$ contributes to the increase of
the pion mass. Also, for large $p$, the dispersion curves approach the
light cone, always from within the causal region $p_0^2 > p^2$.

\begin{figure} 
{\centering
\resizebox*{0.4\textwidth}{0.2\textheight}{\includegraphics{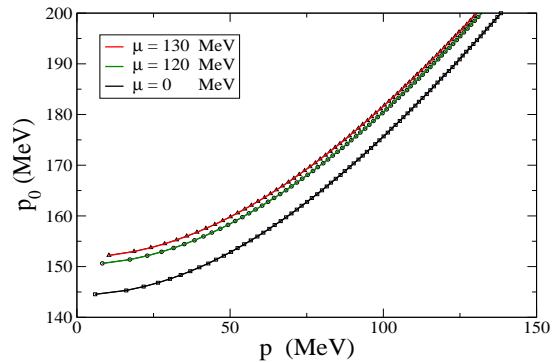}}\par}
\caption{Fit of the dispersion curve to the functional form
$p_0=\sqrt{n^{-1}(T,\mu)p^2 + M^2(T,\mu)}$ for $\mu=0,120,130$
MeV and $T=120$ MeV. The dots are the computed values of the
dispersion relation.} 
\end{figure}

We also notice that the dispersion curves can be parametrized by a
function of the form
\be
   p_0=\sqrt{n^{-1}(T,\mu)p^2 + M^2(T,\mu)}\, .
   \label{parametri}
\ee
This is shown in Fig.~4 where the dots represent the computed values
of the dispersion relation and the continuous curves the
fits. Table~1 shows the values of the parameters $n(T,\mu)$ and 
$M(T,\mu)$ for the different values of $\mu$ and the temperature considered.
From this table we observe that the presence of a finite chemical
potential has two dramatic effects on the behavior of the pion
dispersion curve compared to the case where only the temperature is
considered: first, there is a significant increase in the pion mass
and second, the index of refraction parameter $n$ becomes larger than unity.

Recall that the magnitude of the pion group velocity $v_g$ is given as
the derivative of the pion energy $p_0$ with respect to the momentum 
$p$. From Eq.~(\ref{parametri}) we see that
\be
   v_g&\equiv&\frac{dp_0}{dp}\nonumber\\
   &=&n^{-1}\left(\frac{p}{p_0}\right)\nonumber\\
   &=&n^{-1}\left(\frac{p}{p_0^{\mbox{\tiny vac}}}\right)
       \left(\frac{p_0^{\mbox{\tiny vac}}}{p_0}\right)\nonumber\\
   &=&v_g^{\mbox{\tiny vac}}/\tilde{n}(p)\, ,
   \label{vg}
\ee
where we have defined the (momentum dependent) index of refraction
$\tilde{n}(p)$ as
\be
   \tilde{n}(p)=n\left(\frac{p_0}{p_0^{\mbox{\tiny vac}}}\right)\, ,
   \label{refin}
\ee
and the vacuum pion group velocity $v_g^{\mbox{\tiny vac}}$ as
\be
   v_g^{\mbox{\tiny vac}}=
   \left(\frac{p}{p_0^{\mbox{\tiny vac}}}\right)\, .
   \label{vgvac}
\ee
Equation~(\ref{refin}) can be interpreted as a temperature, density, and
momentum-dependent pion index of refraction. Since for finite $T$ and
$\mu$, we always have both $n>1$ and $p_0/p_0^{\mbox{\tiny vac}}>1$,
for all values of $p$, the index of refraction developed by the pion medium
at finite density and temperature is always larger than unity.   
       
\begin{table}
\begin{center}
\begin{tabular}{|c|c|c|}\hline
   $\mu$ (MeV) & $n$ & $M$ (MeV) \\ \hline
          0    &  1  &   144.411  \\ \hline
         120   &1.004&   150.388  \\ \hline
         130   &1.006&   151.826  \\ \hline
\end{tabular}
\end{center}
\caption{Values of the parameters $n(T,\mu)$ and $M(T,\mu)$, for three
         values of the chemical potential, $\mu=0,120,130$ MeV and for
         a temperature $T=120$ MeV.}
\end{table}

\section{Summary and conclusions}\label{secIV}

In this paper we have considered the effects of a finite pion density
on the pion dispersion curve at finite temperature. The finite density
is described in terms of a finite pion chemical potential. We have argued
that such description is valid during the hadronic phase of a
collision of heavy nuclei at high energies between chemical and
thermal freeze-out when the strong pion-number changing processes have
driven the pion number to a fixed value. 

In order to consider a general scenario that takes into account
resummation effects, we have presented an approximate solution to the
Schwinger-Dyson equation for the momentum-dependent pion
self-energy, writing this as $\Pi(P)\simeq\Pi_0+\tilde{\Pi}(P)$ and
considering $\tilde{\Pi}(P)\ll\Pi_0$, which 
is justified given that in our approximation, $\tilde{\Pi}\sim{\mathcal
O}(\alpha^2)$, whereas the perturbative expansion of $\Pi_0$ starts at
${\mathcal O}(\alpha)$.
 
The pion dispersion relation thus obtained at finite density and
temperature deviates from the vacuum dispersion relation and can be
described in terms of a density and temperature dependent mass and an
index of refraction, also temperature, density as well as momentum
dependent. This index of refraction is larger than unity for all
values of the momentum. Similar results have been obtained in
Ref.~\cite{Pisarski} using also a linear sigma model {\it without} the
introduction of a finite chemical potential but rather as a
consequence of the loss of Lorentz invariance when a particle travels in a
medium at finite temperature. We note however that our result is more
general than that of the former reference where the computation of the
pion dispersion relation was carried from the onset in the weak coupling
regime, that is to say, $\lambda^2 \ll 1$, whereas our approach is
valid for arbitrary values of $\lambda^2$~\cite{Ayala}.

A very interesting consequence of the development of such index of
refraction happens when considering pions that approach the boundary
between the strongly interacting hadronic region of the collision and
vacuum. Propagation through the boundary {\it wall} can be described in three
different regimes, depending on whether the wall's width $d$ is much
smaller, on the order of, or much larger than the pion's wave length
$\lambda$. In the first situation (thin wall regime), it is possible
to treat the propagation of the pions through the boundary in terms of
transmission and reflection coefficients. In the second case,
diffusion effects have to be properly taken into account. However, in
the third case ($d \ll \lambda$), which we choose here as an
illustrative scenario, it is possible to consider the motion of pions
through the wall as that of classical particles climbing out of a
potential well and the interaction of the wall on the pions as that of
a classical force that causes the momentum component normal to the
boundary, $p_t$, to change, while the momentum component parallel to
the boundary, $p_l$, is left unchanged~\cite{Shuryak}. The change is
found by imposing energy conservation for the waves on both sides of
he wall
\be
   \sqrt{n^{-1}(T,\mu)p_{\mbox {\tiny in}}^2+M^2(T,\mu)}=
   \sqrt{p_{\mbox {\tiny out}}^2+m_\pi^2}
   \label{conser1}
\ee
which in turn implies
\be
   (p_t^2)_{\mbox {\tiny out}}=n^{-1}(p_t^2)_{\mbox {\tiny in}}
   -p_l^2[1-n^{-1}] + [M^2-m_\pi^2]\, .
   \label{conser2}
\ee
Notice that when $n=1$, Eq.~(\ref{conser2}) means that the momentum
component normal to the boundary in vacuum grows when compared to the
same component inside the pion medium. However, in the general case of
finite $T$ and $\mu$, $n^{-1}<1$ so, from Eq.~(\ref{conser2}) we see
that it is possible that for certain values of $p_l$,
$p_l^2[1-n^{-1}]\geq n^{-1}(p_t^2)_{\mbox {\tiny in}} + [M^2-m_\pi^2]$
and thus that $(p_t)_{\mbox {\tiny out}}$ be zero or purely imaginary,
meaning that the pion is reflected back.  

These phenomena might have a word on the behavior of the transverse
pion spectra measured from SIS (Heavy Ion Synchrotron) to RHIC 
energies, but clearly a more  
detailed study of the propagation properties of pions through the
boundary between the hadronic phase and vacuum in relativistic
heavy-ion collisions is called for. This will be the subject of a
future work~\cite{progress}.

\section*{Acknowledgments}

A. Aranda and A. Ayala wish to thank Universidad de Colima for
their kind hospitality during the time when part of this work was done.
Support for this work has been received in part by DGAPA-UNAM under PAPIIT
grant number IN108001, CONACyT under grant number 35792-E and 
No. 32279-E, the Department of Energy under grant DE-FG02-91ER40676 
and by Fondo Alvarez-Buylla, Universidad de Colima.

\end{document}